# THERMAL MEASUREMENT AND MODELING OF MULTI-DIE PACKAGES

*András Poppe[1,2], Yan Zhang[3], John Wilson[3], Gábor Farkas[2], Péter Szabó[2], John Parry[4]*
[1]Budapest University of Technology, Department of Electron Devices, Hungary
[2]MicReD Ltd, Hungary, [3]Flomerics Inc., USA, [4]Flomerics Ltd., UK

**ABSTRACT**

Thermal measurement and modeling of multi-die packages with vertical (stacked) and lateral arrangement became a hot topic recently in different fields like RAM chip packaging or LEDs and LED assemblies. In our present study we present results for a more complex structure: an opto-coupler device with 4 chips in a combined lateral and vertical arrangement. The paper gives an overview of measurement and modeling techniques and results for stacked and MCM structures. It describes actual measurement results along with our structure function based methodology which helps validating the detailed model of the package being studied. Also, we show how one can derive junction-to-pin thermal resistances with a technique using structure functions.

## 1. INTRODUCTION

Thermal measurement and modeling of multi-die packages became a hot topic in recent years. A detailed, comprehensive overview has been given recently [1] where different measurement and modeling techniques have been referred to. Nowadays, besides IR techniques (e.g. [2]) the mainstream characterization technique seems to be the JEDEC JESD51-1 based electrical test method [3], yielding either *thermal resistance* values only (static characterization – see e.g. [4], [5]) or providing the *dynamic description* of the packaged multi-die system by means of thermal impedances in various forms (full set of *real heating or cooling curves*, *complex loci* – see e.g. [6] or *structure functions* as presented in [7]).

In many cases thermal transient measurements are used to derive steady-state metrics for multi-die systems [7], [8] but there is no agreement yet on what these metrics should be. One approach is to try to derive a single $R_{th}$ value to represent and model a multi die package [2], [4]. A next step in representing multi die packages is to use multiple thermal resistances [5], [7]. A recent tendency is to measure the temperature change on all chips in the package and to report the results of all these measurements. In what format and with what content this reporting should be done, is still open to discussion [9], but measuring and reporting a full thermal *resistance matrix* [7], [8] or *thermal impedance matrix* [6], [10] seems to gain wider acceptance among different thermal research teams. The attempts to create thermal models out of measurement results include resistor networks with a few elements only [5], [7] or providing the network representation of the complete thermal resistance matrix [8] or providing all the elements of the thermal impedance matrix by means of time-domain or frequency-domain functions [6], [10] or even by means of a dynamic compact model derived from structure functions [10].

In section 2 we provide an overview of thermal transient measurement based characterization through a few typical examples. In section 3 we introduce our recent results in modeling of single and multi die packages with a special emphasis on structure function based detailed model verification. Section 4 presents a case study about an opto-coupler device including four chips both in lateral and vertical arrangement. Through this case study we also present our novel technique to obtain junction-to-pin thermal resistance values from structure functions.

## 2. OVERVIEW OF CHARACTERIZATION OF TYPICAL MULTI DIE PACKAGES

Typical multi die packages contain dies either in a vertical (stacked) arrangement or in a lateral arrangement.

As the silicon technology continues to obey Moore's law – according to which, the number of IC elements on a unit silicon area doubles every 18-24 months – layering the silicon chips on top of each other within a package multiplies the increase originated by shrinking transistor size, by the number of layered dice. 3D stacked die packages are especially common today in RAM packaging and in hand-held devices, especially in cell phones and digital cameras, which require fast turnaround, very high level of integration and low cost that is characteristic in general for System-in-Package (SiP) solutions. Another typical application is integrating chips realized by different technologies into a single package.

In our first example we present a stacked arrangement. A 144LQFP package containing two test dies (cross-sectional view in Figure 1) has been characterized both in JEDEC standard still-air environment and in a cold-plate setup [10]. In each test environment thermal transient measurements have been carried out. A power step has been applied on the top, and later, on the bottom die in a sequential manner and the transient responses were captured in each case on both dies. This way, for each test environment we obtained all the elements of the package's *thermal impedance matrix*: two *driving-point impedances* (heating and measurement at the same location) and two transfer impedances (heating the top die and capturing the temperature on the bottom die and vice versa).

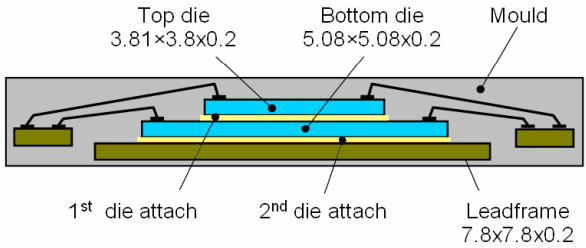

**Figure 1: Two test dies stacked in a 144 LQFP package.**

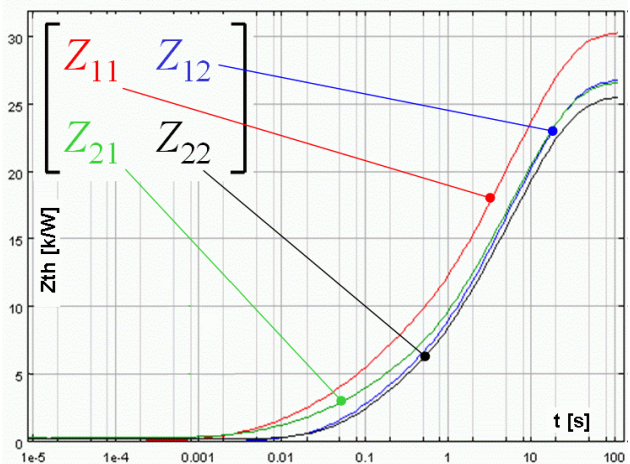

**Figure 2: The thermal impedance matrix of the stacked die package of Figure 1. Matrix elements are represented by time-domain transient impedance (Zth) curves.**

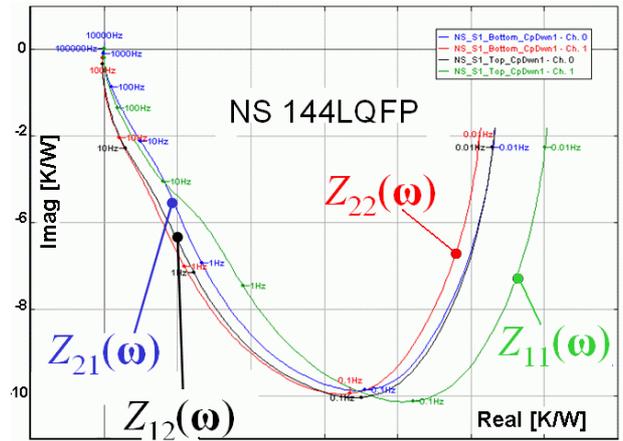

**Figure 3: Frequency-domain representation of the thermal impedance matrix of the stacked die package**

Figure 2 presents the measurement results for a cold-plate environment. As one can see, the $Z_{th}$ thermal impedance matrix of the package shows asymmetry. $Z_{12}$ and $Z_{21}$ elements of the matrix (thermal transfer impedances between the top and bottom dies) differ in the small time-constant range suggesting, that the top-to-bottom and bottom-to-top heat transfer differ, due to the different size of the dies.

As an alternate representation, the $a_{lk}(t)$ time-domain response of the *l*-th die when a unit power step is applied at die *k* can be transformed into the frequency-domain:

$$Z_{lk}(\omega) = \frac{1}{p}\int_0^\infty a_{lk}(t)e^{-j\omega t}dt \quad . \tag{1}$$

Figure 3 shows the *frequency-domain representation of the impedance matrix* of the 144LQFP package. Again, the asymmetry can be observed. This asymmetry in the impedance matrix means *non-reciprocal behavior* which, when a compact model is to be constructed, has to be accounted for.

As a next example, we show a lateral arrangement of four power DMOS switches in a P-TO263-15-1 package. The center chip contains two switches, two other chips on separate tabs have single switches [6].

This package (Figure 4) has also been characterized with thermal transient measurements: all elements of its thermal impedance matrix have been measured, both in a still-air and in a cold-plate setup.

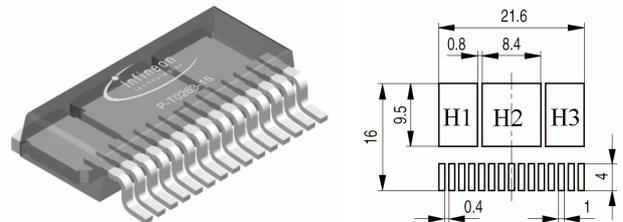

**Figure 4: P-TO263-15-1 package, internal leadframes, footprint and tab numbering (H1, H2, H3).**

Measuring the package in still-air setup we got the frequency-domain representation of the impedance matrix plotted in Figure 5. Driving the larger chip on the H2 center tab the self-impedance ($Z_{22}$) is lower than when driving the smaller chip ($Z_{11}$). Again, the off-diagonal elements show non-reciprocal behavior, $Z_{12} \neq Z_{21}$.

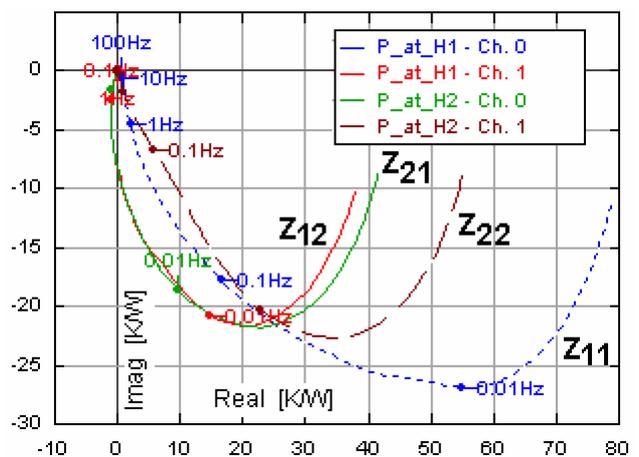

**Figure 5: Comparison of complex loci in still-air setup. Curves $Z_{11}$, $Z_{12}$ measured with junction on H1 driven, $Z_{22}$, $Z_{21}$ measured with junction on H2 driven.**

In general, one can say, that in a given test environment a multi die package (having any kind of arrangement) is totally represented by its full thermal impedance matrix in the form shown by Eq. 2. Here $Z_{lk}$ off-diagonal elements represent thermal transfer impedances between junctions $l$ and $k$, $Z_{ii}$ describes the self-impedance or driving point impedance at the junction of the $i$-th die.

The impedance matrix elements can be described either by time-domain functions (impedance curves) or by frequency-domain functions (e.g. by complex loci) or by network models of the impedances. Multiplying this matrix with the vector of any combination of $p_i$ powers applied at the chips, one can obtain the corresponding vector of the $\tau_i$ temperature responses at all dies.

$$\begin{bmatrix} \tau_1 \\ \tau_2 \\ \vdots \\ \tau_n \end{bmatrix} = \begin{bmatrix} Z_{11} & Z_{12} & \cdots & \\ Z_{21} & & & \\ \vdots & & & \vdots \\ Z_{n1} & & \cdots & Z_{nn} \end{bmatrix} \begin{bmatrix} p_1 \\ p_2 \\ \vdots \\ p_n \end{bmatrix} \quad (2)$$

*Note*, that the above impedance matrix reduces to a thermal resistance matrix if the steady-state values of the time-domain impedances (t=∞) or if the 0 Hz value of the frequency-domain impedance values are considered. Such an $R_{th}$ matrix is presented for multi-die LED assembly in [8] where asymmetry of the off-diagonal elements of the matrix has also been reported.

With $N$ chips in a package, there are $N^2$ thermal impedances present. $N$ driving point impedances describe the heat-removal properties from the junction on a die towards the ambient. For every die there are $N-1$ other impedances that describe the properties of heat transfer from the driven die to any other die in the package.

The matrix elements can be identified by $N$ thermal transient measurements. In measurement $i$, the $i$-th die is excited (a power step is applied to it) and the temperature responses on all dies are measured and recorded. The measurements can be carried out in standard test environments as prescribed by the JEDEC JESD51 series of standards [3]. In this way the *thermal impedance matrix concept* is a natural, unambiguous *extension of the JEDEC JESD51-1 concept of a single junction-to-ambient resistance* for multi die packages. It describes steady-state properties and dynamic behavior of a multi die package.

### 3. MODELING ISSUES

#### 3.1. Compact modeling

As mentioned in the introduction, there have been several attempts to develop compact models of multi die packages. Many teams have aimed to create steady-state models by means of a single or just a few thermal resistances [2], [4], [5], [7] or by means of a complete thermal resistance matrix [8]. Our team has focused its efforts on developing general models. In one approach to modeling multi die packages, the matrix representation of the package is used with the aim of developing compact models which also exhibit the reported *non-reciprocal behavior*. Such a model is shown in Figure 6.

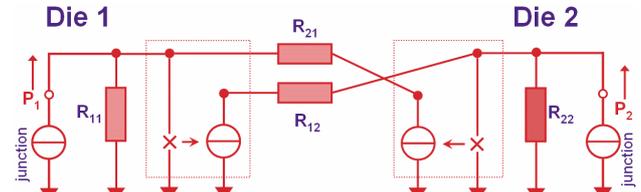

**Figure 6: Non-reciprocal steady-state compact model of a two-die package.**

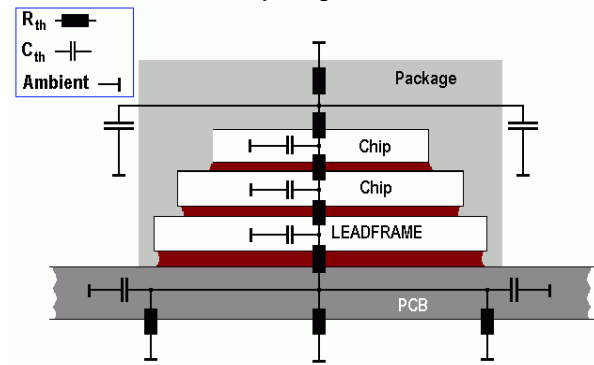

**Figure 7: A dynamic compact model of stacked die package.**

The resistor values are to be taken from the thermal resistance matrix of the package: $R_{11}$ and $R_{22}$ (elements of the main diagonal) describe the self-heating when die 1 and die 2 are heated, respectively. Resistors $R_{12}$ and $R_{21}$ describe the coupling between the two dies. The temperature controlled heat-flux generators (framed in the figure) ensure that either $R_{12}$ or $R_{21}$ is effective, depending at which die heating is applied, allowing the entire model to reflect the non-reciprocity (asymmetry) shown by the measurement results (see previous section).

If dynamic behavior is to be considered the element values of the impedance matrix need to be represented by a proper approximation (e.g. with more RC elements) and should be used in the model shown in Figure 6.

Another compact modeling approach tries to derive dynamic compact models from measurement results using *structure functions*. Such a model for a stacked die package is shown in Figure 7. Element values for this model can be obtained from a step-wise approximation of the *cumulative structure function* corresponding to the driving point thermal impedance measured at the top die [1], [10].

Boundary condition independent compact models (either steady-state or dynamic) have not been widely reported for multi die packages. At present our team carries out research in this direction. Our approach tries to extend the DELPHI methodology [11], [12] in two directions. One goal is to properly account for multiple dies and the possibly asymmetrical couplings among them. Another goal is to create dynamic models.

### 3.2. Structure functions for validation of detailed models

In the DELPHI methodology and its extension to single die transient package models creation of boundary condition independent compact models is based on validated detailed models. For this purpose thermal measurements are carried out in four different dual cold plate setups [13] and results are compared against simulation results. Now, instead of comparing the raw transient curves themselves we suggest using *structure functions for model validation*.

The reason behind this is that if a detailed model properly describes physical reality, the structure functions derived from simulated transient responses should show exactly the same features as the structure functions derived from the thermal transient measurements of the real physical package. The suggested flow of model validation is shown in Figure 8.

This model validation method is of special interest if the detailed model is to be used for generating *dynamic compact models* since here the distribution of thermal capacitances and thermal resistances along a given major heat-flow path is essential.

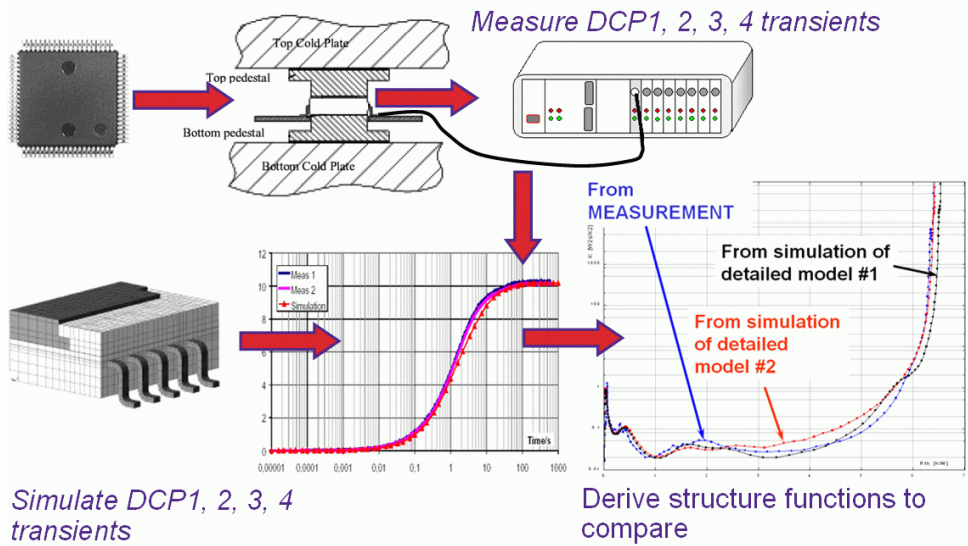

**Figure 8:** Suggested flow of validation of detailed models with the help of structure functions.

In case of modeling multi-die packages, especially with stacked structures, the proper modeling of the *internal parts of the package* is important. For a stacked die arrangement, the chip-to-chip couplings have to be properly described, for which we should fine tune the detailed models of the various chips and die attach layers. With the help of the structure functions one should be able to differentiate between the dies and optimize the model of the die attach layers.

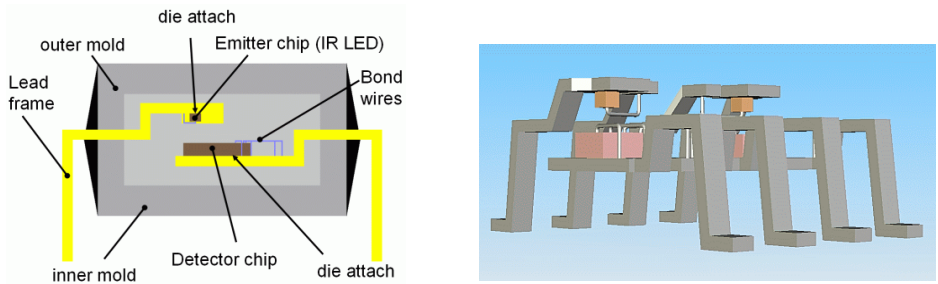

**Figure 9:** A dual opto-coupler devices in a plastic DIL package containing four chips: cross-sectional view of the package and 3D axonometric view of the detailed CAD model

### 4. A CASE STUDY

In this study we present thermal transient measurement and simulation results of dual opto-coupler devices. The plastic DIL package contained four chips. A single opto-coupler had its two chips in a vertical arrangement: there was an IR LED (*emitter*) on the top and a photo-transistor (*detector*) underneath. The package had two leadframes: one containing the emitter chips, another the detector chips in a lateral arrangement. We had two versions of the opto-coupler: with a low power detector and with a detector having a power output stage. The general geometry of the studied devices is shown in Figure 9.

We had our samples attached to JEDEC standard thermal test boards with low and high thermal conductivity which were measured in a JEDEC standard 1 ft$^3$ still-air chamber and we measured a few samples also in a cold plate environment.

### 4.1. Test results

In our study we were interested in the driving point impedances of the *emitter* and *detector* chips as well as in the *emitter-detector*, *emitter-emitter* and *detector-detector* thermal couplings. We derived JEDEC standard thermal metrics from structure functions which were obtained by thermal transient test using the *T3Ster* equipment. First we validated the detailed model of the package using the FLOTHERM program and using cold plate measurements. In a next step this validated model was attached to the model of the still-air environment. In this way we were able to derive:

- junction-to-ambient thermal resistance values for the emitter and detector chips,
- junction-to-case and
- junction-to-pin (junction-to-lead) thermal resistance values for both types of chips,
- the transfer conductances between the various chips inside the packages,
- and the validated model of the packages,

all from thermal transient measurements using structure functions. The term "junction-to-case" here denotes simply the partial thermal resistance between the junction and the bottom of the package, which has actually no "case", i.e. exposed tab. We introduced this measurement only to increase the number of boundaries for modeling.

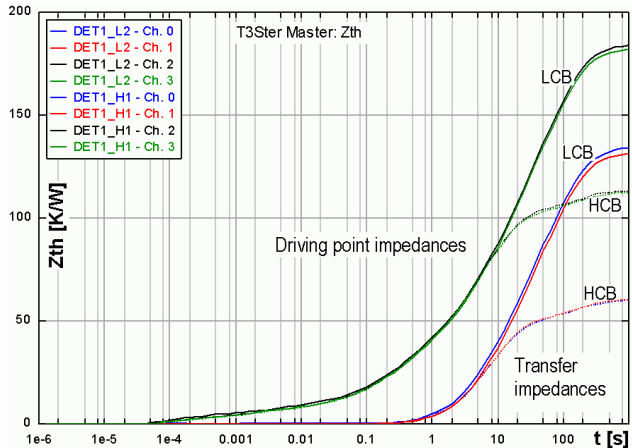

**Figure 10:** Measured thermal impedances with detector driven in "low power" samples. Still-air environment, low conductance (LCB) and high conductance (HCB) board.

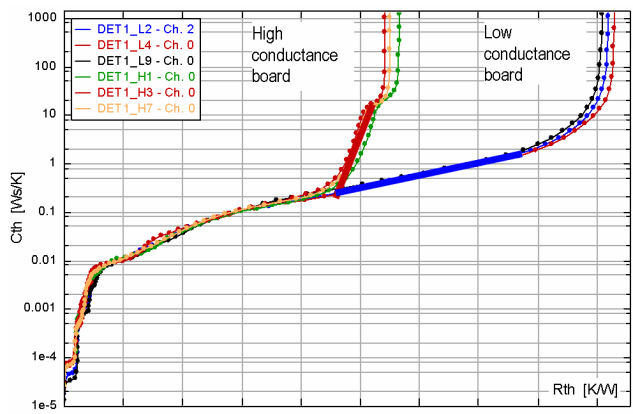

**Figure 11:** Cumulative structure functions of driving point thermal impedances in the same arrangement.

In Figure 10 we present some measurement results for the first type with low power detector, Figure 11 shows structure functions. In this figure the straight thick lines added to the plot correspond to the radial heat-spreading in the test boards. So, as suggested already by $Z_{th}$ curves one can distinguish the heat-flow inside the package and in the test environment. It is worth noting that all structure functions coincide in the regions describing the heat-flow in the package and pins, whatever test board is used. Note also that the test environment represents a considerable part of the total junction-to-ambient thermal resistance, in case of the low conductivity board this is almost 50%.

**4.2. Steady-state thermal metrics**

JEDEC type steady-state thermal metrics have been identified both from the physical test results and from the simulation of the validated detailed models. *The junction-to-ambient thermal resistance values* are read directly from the structure functions: the location of the singularity provides the $R_{thJA}$ value for any test environment. For the identification of *junction-to-case thermal resistance values* the so-called *dual interface method* was used [6]. Here two measurements were carried out; one with the "case" surface of the package in a direct contact with a cold plate and a second with a thermal insulator layer inserted between the package and the cold plate. The resulting structure functions are shown in Figure 12.

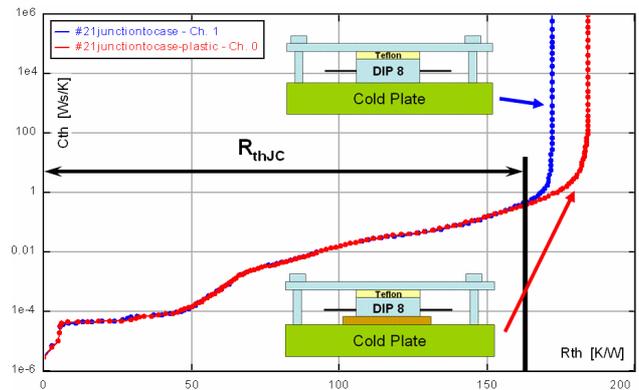

**Figure 12:** Identification of $R_{thJC}$ values for the second series of devices ($R_{thJC}$ for the detector chip).

The *junction-to-pin thermal resistance values* were identified using a similar method:

- The samples were measured in a JEDEC standard still-air environment,
- The measurements were repeated with an extra thermal mass attached to the pins. Structure functions were derived from both sets of thermal transient curves.
- As the structural change occurred at the pins, thus the location where deviation is observed in the structure functions gives the value of the junction-to-pin thermal resistance (Figure 13).

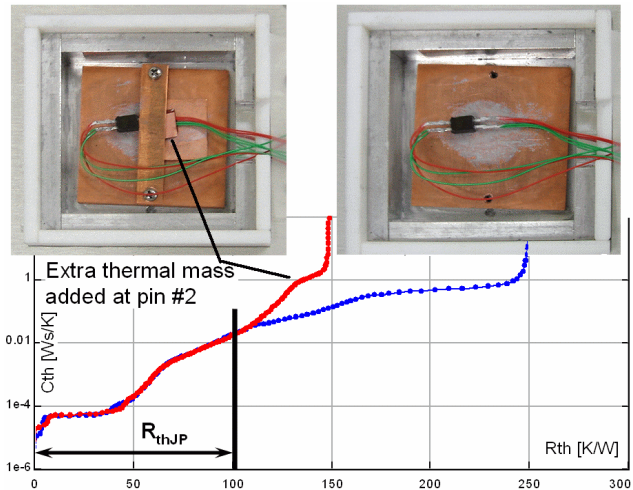

**Figure 13:** Cold-plate setups for the identification of the $R_{thJP}$ value for the "high power" type of devices ($R_{thJP}$ for the detector chip)

## 5. CONCLUSIONS

In this paper we gave an overview on measurement and modeling of multi die packages. Packages with both vertical and lateral chip arrangements have been discussed.

Several measurement examples have demonstrated that coupling between different chips in multi-die packages may show asymmetry, which has also been reported by other research teams. This asymmetry necessitates the use of network elements which have been unusual in compact thermal models. *Temperature controlled heat-flux generators* can be used to properly model the non-reciprocal behavior observed in measurements.

The use of thermal resistance matrices (steady-state description) or *thermal impedance matrices* (dynamic case) provides an unambiguous *extension of usual single-die thermal metrics to multi-die packages*.

Besides steady-state compact models using elements of thermal resistance matrices, we outlined dynamic compact models of stacked die packages derived from structure functions.

When simulation of the detailed model does not fit measurements in the physical structure then structure functions help identify parts where deviation occurs. This validation of detailed models is of primary importance for stacked die packages.

Finally we presented a case study with four dies both in a vertical and lateral arrangement. The elements of the thermal impedance matrix of that package were identified in the form of thermal impedance curves for different types of test environments. Structure function based methods were used to obtain some steady-state thermal metrics of the package: the modification of the dual interface method was suggested to derive junction-to-pin thermal resistances. Validated by test results, the detailed model of the package has been also used to derive thermal metrics in other standard test environments.